\documentclass[10pt,prl,preprintnumbers,floatfix,aps,nofootinbib,showpacs,twocolumn,superscriptaddress]{revtex4} 

\usepackage{epsfig}
\usepackage{url}
\usepackage{hyperref}

\usepackage[normalem]{ulem} 

\usepackage{latexsym}
\usepackage{epsfig}
\usepackage{amsmath}
\usepackage{amssymb}
\usepackage{wasysym}
\usepackage{graphicx}
\usepackage{verbatim}
\usepackage{enumerate,mdwlist}
\usepackage[titletoc]{appendix}
\usepackage{amsfonts}
\usepackage{tikz} 
\usetikzlibrary{calc}
\usepackage{pgfplots}
\usepackage[export]{adjustbox}

\usepackage[normalem]{ulem}

\newcommand{\lp}{\left(}
\newcommand{\rp}{\right)}

\newcommand{\ba}{\begin{eqnarray}}
\newcommand{\ea}{\end{eqnarray}}
\newcommand{\be}{\begin{equation}}
\newcommand{\ee}{\end{equation}}

\newcommand{\al}{\alpha}

\newcommand{\Lag}{\mathcal{L}}

\newcommand{\aT}{\alpha_{_T}}

\newcommand{\aM}{\alpha_{_M}}
\newcommand{\Ms}{M_{*}}

\newcommand{\dl}{d_{_L}} 

\definecolor{grey}{rgb}{0.4,0.4,0.4}
\definecolor{dullmagenta}{rgb}{0.4,0,0.4}
\definecolor{darkblue}{rgb}{0,0,0.4}
\definecolor{midblue}{rgb}{0,0,0.5}
\definecolor{midred}{rgb}{0.5,0,0}
\definecolor{orange}{rgb}{1,0.5,0}
\definecolor{lightbrown}{rgb}{0.75,0.5,0.25}
\definecolor{tan}{cmyk}{0.14,0.42,0.56,0}
\definecolor{djunglegreen}{cmyk}{0.99,0,0.52,0}
\definecolor{lightgreen}{rgb}{0,1,0}
\definecolor{olivegreen}{cmyk}{0.64,0,0.95,0.40}
\definecolor{midgreen}{rgb}{0.0,0.675,0.0}
\definecolor{darkgreen}{rgb}{0,0.5,0}

\usepackage[all]{xy} 
\usepackage{amsfonts}

\begin{document} 

\title{Dark Energy after GW170817: dead ends and the road ahead}

\author{Jose Mar\'ia Ezquiaga}
\email{jose.ezquiaga@uam.es}
\affiliation{Instituto de F\'isica Te\'orica UAM/CSIC, Universidad Aut\'onoma de Madrid, \\ 
C/ Nicol\'as Cabrera 13-15, Cantoblanco, Madrid 28049, Spain}
\affiliation{Berkeley Center for Cosmological Physics, LBNL and University of California at Berkeley, \\
Berkeley, California 94720, USA}

\author{Miguel Zumalac\'arregui}
\email{miguelzuma@berkeley.edu}
\affiliation{Berkeley Center for Cosmological Physics, LBNL and University of California at Berkeley, \\
Berkeley, California 94720, USA}
\affiliation{Nordita, KTH Royal Institute of Technology and Stockholm University, \\
Roslagstullsbacken 23, SE-106 91 Stockholm, Sweden}
\affiliation{Institut de Physique Th\' eorique, Universit\'e  Paris Saclay 
CEA, CNRS, 91191 Gif-sur-Yvette, France}

\begin{abstract}
Multi-messenger gravitational wave (GW) astronomy has commenced with the detection of the binary neutron star merger GW170817 and its associated electromagnetic counterparts. The almost coincident observation of both signals places an exquisite bound on the GW speed $|c_g/c-1|\leq5\cdot10^{-16}$. We use this result to probe the nature of dark energy (DE), showing that a large class of scalar-tensor theories and DE models are highly disfavored. As an example we consider the covariant Galileon, a cosmologically viable, well motivated gravity theory which predicts a variable GW speed at low redshift. Our results eliminate any late-universe application of these models, as well as their Horndeski and most of their beyond Horndeski generalizations.
Three alternatives (and their combinations) emerge as the only possible scalar-tensor DE models: 1) restricting Horndeski's action to its simplest terms, 2) applying a conformal transformation which preserves the causal structure and 3) compensating the different terms that modify the GW speed (to be robust, the compensation has to be independent on the background on which GWs propagate).
Our conclusions extend to any other gravity theory predicting varying $c_g$ such as Einstein-Aether, Ho\v{r}ava gravity, Generalized Proca, TeVeS and other MOND-like gravities.
\end{abstract}

\date{\today}


 
\maketitle
 
\paragraph{\bf Probing Dark Energy with GWs.}

Multi-messenger gravitational wave (GW) astronomy became a reality with the detection of a binary neutron star (BNS) merger with GWs by LIGO-VIRGO collaboration (GW170817) \cite{DiscoveryPaper} and subsequently with different electromagnetic (EM) counterparts by Fermi \cite{2041-8205-848-2-L13} and a range of observatories accross the spectrum \cite{2041-8205-848-2-L12}. This extraordinary discovery has many potential applications to test the astrophysics of BNS mergers \cite{Faber:2012rw}, the fundamentals of gravity in the strong regime \cite{Wex2016} and cosmic expansion \cite{Nissanke:2013fka}. In this letter we present the implications that this measurement has for the nature of dark energy (DE) and tests of General Relativity (GR).

The present cosmic acceleration is probably one of the greatest challenges in modern physics. Leaving the theoretical fine tuning issues aside \cite{Weinberg:1988cp}, a cosmological constant is the leading candidate to explain this acceleration since it is fully consistent with observations \cite{Ade:2015xua}. Alternative scenarios that explain DE dynamically require either additional degrees of freedom (beyond the massless spin-2 field of GR) or a low-energy violation of fundamental principles such as locality \cite{Clifton:2011jh}. The extremely low energy scale for DE requires additional degrees of freedom to be hidden on small scales by a screening mechanism \cite{Joyce:2014kja}, which also suppresses their rate of emission as additional gravitational wave polarizations \cite{deRham:2012fw}.

New fields coupled to gravity can affect the propagation speed of the standard GW polarizations, as measured by GW170817 and its counterparts \cite{Bettoni:2016mij}. Anomalous GW speed can be used to test even screened theories, as signals from extra-galactic sources probe unscreened, cosmological scales. In addition, effects on GW propagation accumulate over the travel time of the signals, amplifying their magnitude and yielding an impressive sensitivity. GW astronomy is therefore the most powerful tool to test models that modify GW propagation.

Some of the most interesting dark energy models predict an anomalous GW speed and are ruled out by GW170817. These include cosmologically viable, screened and self-accelerating models like the covariant Galileon \cite{Nicolis:2008in,Deffayet:2009wt}, or proposals to solve the cosmological constant problem like the Fab-four \cite{Charmousis:2011bf}. We will describe the implications of GW170817 on these and other DE models, determining which of them remain viable after this discovery. We will focus on gravity theories with just one additional mode, a scalar field, working in the framework of Horndeski \cite{Horndeski:1974wa} and beyond Horndeski \cite{Zumalacarregui:2013pma} GLPV \cite{Gleyzes:2014dya,Deffayet:2015qwa} and Degenerate Higher-Order Scalar-Tensor (DHOST) \cite{Langlois:2015cwa,Crisostomi:2016czh,Achour:2016rkg,BenAchour:2016fzp} theories. Nevertheless, our analysis can be extended to theories with more degrees of freedom such as massive gravity \cite{deRham:2010kj}, Einstein-Aether theories \cite{Jacobson:2004ts}, Ho\v{r}ava gravity \cite{Blas:2010hb} or TeVeS \cite{Sagi:2010ei}.

\paragraph{\bf GW170817 and its counterparts.}
\label{sec:LIGOresults}

On August 17, 2017 the LIGO-VIRGO collaboration detected the first BNS merger, GW170817 \cite{DiscoveryPaper}. This event was followed-up by a short gamma ray burst (sGRB), GRB170817A, seen just $1.74\pm0.05$s later by Fermi and the International Gamma-Ray Astrophysics Laboratory \cite{2041-8205-848-2-L13}. Subsequent observations across the electromagnetic spectrum further confirmed the discovery \cite{2041-8205-848-2-L12}.

Each of these events provides complementary information about the BNS merger. The GW signal serves to weight the NS, which are in the range $0.86-2.26 M_\odot$, and to measure the luminosity distance, $\dl=40^{+8}_{-14}$Mpc. The EM counterparts uniquely identify the host galaxy, NGC4993. 
Taking the lowest limit $\dl=26$Mpc and a conservative $10$s delay between the GW and sGRB the bound on the speed of GWs is \cite{2041-8205-848-2-L13}
 \be \label{eq:bound_speed}
 -3\cdot 10^{-15} \leq c_g/c-1\leq 7\cdot 10^{-16}\,.
 \ee
This is many orders of magnitude more stringent than previous direct bounds \cite{Cornish:2017jml} and applies to $c_g>c$ unlike bounds from absence of gravitational Cherenkov radiation \cite{Moore:2001bv}. For simplicity, we will use a symmetric bound $|c_g/c-1|\leq 5\cdot 10^{-16}$ in the rest of the letter. Hereafter we use natural units with $c=1$.

\paragraph{\bf GW propagation in scalar-tensor gravity.}\label{sec:propagation_effects}

Effects on the propagation of GWs are a hallmark of scalar-tensor theories of gravity.
The evolution of  linear, transverse-traceless perturbations over a cosmological background
\begin{equation}
  \label{eq:cosmo_tensors}
\ddot{h}_{ij}+(3+\aM)H\dot{h}_{ij}+(1+\aT)k^2h_{ij}=0\,,
\end{equation}
is fully characterized by two functions of time: the \emph{tensor speed excess}, $\aT$, which modifies the propagation speed of GWs $c_g^2=1+\aT$ and hence the causal structure for this type of signal; and the \emph{running of the effective Planck mass}, $\aM\equiv d \log(M_*^2)/d\log(a)$, which modulates the friction term caused by the universe's expansion. 
These functions depend on the theory parameters and the cosmological dynamics of the scalar field. The explicit expressions are given for 
Horndeski in \cite{Bellini:2014fua}, for beyond Horndeski GLPV in \cite{Gleyzes:2014qga} and for DHOST theories in \cite{Langlois:2017mxy}. The constraint on $c_g$ (\ref{eq:bound_speed}) has fundamental implications for DE scenarios and can by itself rule out otherwise viable models, as we will see explicitly now for the Covariant Galileon.

\paragraph{\bf The fate of covariant Galileon.}\label{sec:galileon}

\begin{figure*}[t!]
\centering
\includegraphics[width = 0.50\textwidth,valign=t]{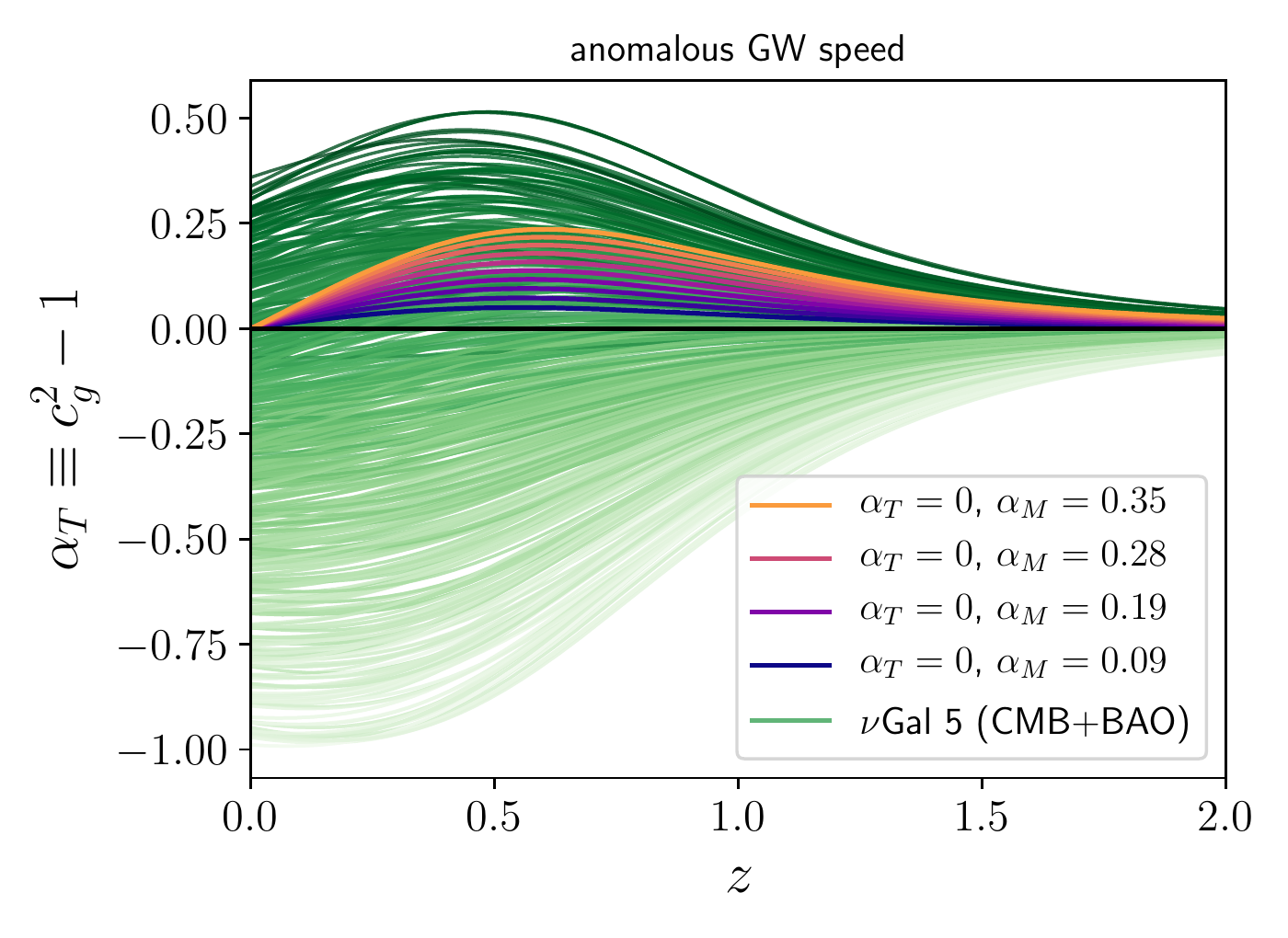}
\includegraphics[width = 0.48\textwidth,valign=t]{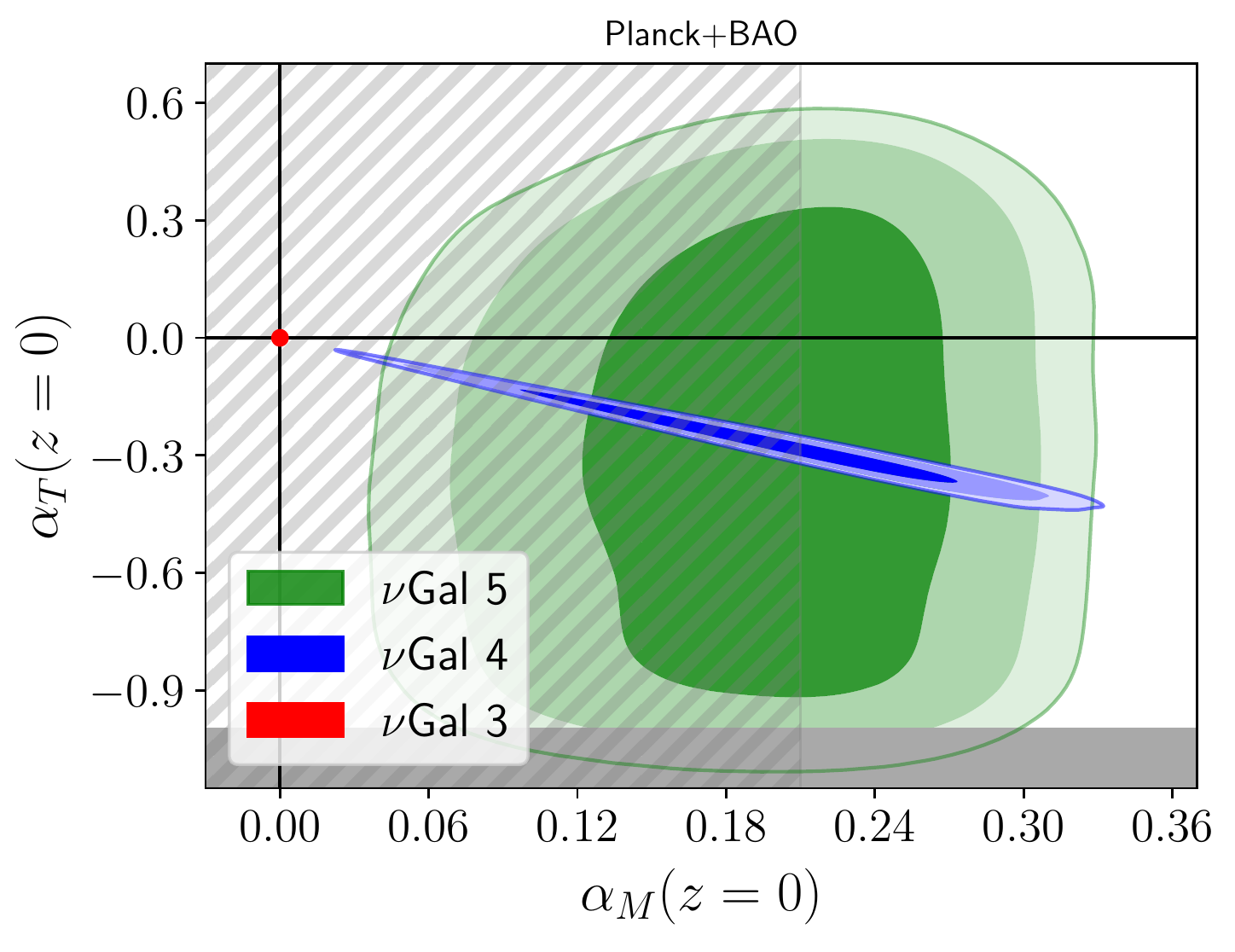}
\caption{\textbf{Left:} time evolution of the tensor speed excess $\aT$ as a function of redshift for 300 different realizations of viable quintic Galileon cosmologies. Only quintic fine tuned cases (colored) predict $\aT(z=0)\approx0$. \textbf{Right:} 1, 2 and 3$\sigma$ confidence regions of the parameter space w.r.t. Planck+BAO for cubic (red), quartic (blue) and quintic (green) Galileons, projected on the $\aT(z=0),\aM(z=0)$ plane. Gray diagonal lines indicate the region disfavored by CMB-LSS cross correlation, measuring the ISW effect (see \cite{Renk:2017rzu} for details). Models with $\aT<-1$ (gray filled region) have unstable tensor modes.}
\label{fig:aT}
\end{figure*}

Galileon gravity is an interesting example of a dark energy model that can be thoroughly tested by GW observations.
It arises from a scalar field with non-linear derivative self-interactions satisfying the Galilean symmetry $\phi\to\phi + C + b_\mu x^\mu$ in flat space-time \cite{Nicolis:2008in}.
Its covariant generalization \cite{Deffayet:2009wt,Deffayet:2011gz} is a simple instance of Horndeski's theory \cite{Horndeski:1974wa}, whose action reads \cite{Kobayashi:2011nu}
\begin{equation}\label{eq:Li}
S[g_{\mu\nu},\phi]=\int\mathrm{d}^{4}x\,\sqrt{-g}\left[\sum_{i=2}^{5}{\cal L}_{i}
\,+\mathcal{L}_{\text{m}}
\right]\,,
\end{equation}
with
\begin{eqnarray}
{\cal L}_{2} & = & G_{2}(\phi,X) \,,
\qquad\qquad {\cal L}_{3} \; = \; G_{3}(\phi,X)\Box\phi\,,\label{eq:L3}\\
{\cal L}_{4} & = & G_{4}(\phi,X)R
+G_{4,X}(\phi,X)\left[\left(\Box\phi\right)^{2}-\phi_{;\mu\nu}\phi^{;\mu\nu}\right]\,,\label{eq:L4}\\
{\cal L}_{5} & = & G_{5}G_{\mu\nu}\phi^{;\mu\nu}
-\frac{1}{6}G_{5,X}(\phi,X)\Big[\left(\Box\phi\right)^{3}
\nonumber \\ && \qquad\qquad
-3\phi_{;\mu\nu}\phi^{;\mu\nu}\Box\phi
+2{\phi_{;\mu}}^{\nu}{\phi_{;\nu}}^{\alpha}{\phi_{;\alpha}}^{\mu}\Big]\,.\label{eq:L5}
\end{eqnarray}
The covariant Galileon corresponds to
\be
\begin{split}&G_{2}(X)=c_{2}X 
\,, \\ &G_{4}(X)=\frac{M_{p}^{2}}{2}+\frac{c_{4}}{M^{6}}X^{2}\,, \end{split} \qquad 
\begin{split}&G_{3}(X)=2\frac{c_{3}}{M^{3}}X \,, \\&G_{5}(X)=\frac{c_{5}}{M^{9}}X^{2} \,,\end{split}
\ee
so that all the coefficients of the second-derivative terms are proportional to $X$. Here $g$ is the determinant of the metric $g_{\mu\nu}$, $R$ is the Ricci scalar, $G_{\mu\nu}$ is the Einstein tensor, $X\equiv-\partial_{\mu}\phi\partial^{\mu}\phi/2$, $\phi_{;\mu\nu} = \nabla_\mu \nabla_\nu\phi$, $\Box\phi = \nabla_\mu\nabla^\mu\phi$ and $\mathcal{L}_{\text m}$ denotes the Lagrangian of some matter field $\psi_{\text m}$. The mass scale $M^3\equiv M_{\rm Pl}H_0^2$ ensures that the $c_i$ coefficients remain dimensionless ($M_{\rm Pl}$ is the Planck mass). 
We will refer to three models depending on the highest power of $\phi$ present in the action (\ref{eq:Li}): \textit{cubic} ($c_4=c_5=0$), \textit{quartic} ($c_5=0$) and \textit{quintic} (all terms). 

The covariant Galileon is most interesting as a cosmological model where the Galileon field causes the universe to self-accelerate (without the need of a cosmological constant). 
As a consequence of shift-symmetry $\phi\to \phi+C$, a tracker solution exists where the time evolution of the field and the Hubble rate obey the relation 
 $\xi \equiv {H(t)\dot\phi(t)}/{H_0^2}=\mathrm{constant}$ 
\cite{DeFelice:2010pv}. Under this solution, which has to be reached before DE domination \cite{Barreira:2013jma}, the functions of the modified GW equation (\ref{eq:cosmo_tensors}) read
\begin{eqnarray}
&&\hspace{-20pt} \aT = \frac{1}{M_*^2E^4}\left[ 2c_4 \xi^4 
+ c_5 \xi^5 \left(1 +  \frac{\dot H}{H^2}\right)\right]\,,  \label{eq:aT_gal}
\\
&&\hspace{-20pt} \aM = -4\frac{\dot H}{H^2}\frac{M_*^2 - 1}{M_*^2} \label{eq:aM_gal}
\,,\quad\!
M_*^2  =1-\frac{\xi^4}{E^4}\left(\frac{3}{2}c_4 + c_5\xi \right),
\end{eqnarray}
where $E = H(t)/H_0$. 

Self-accelerating Galileon models are all consistent (if massive neutrinos are included) with cosmic microwave background (CMB) and baryon acoustic oscillations (BAO), together with the locally measured value of $H_0$ (avoiding the tension in $\Lambda$CDM) \cite{Barreira:2014jha,Renk:2017rzu}. The inclusion of cross-correlations between CMB temperature and galaxies, which probes the Integrated Sachs Wolfe (ISW) effect, trims a significant portion of the parameter space (including all cubic models), but leaves a region that is still viable \cite{Renk:2017rzu}, ($\aM(z=0)\gtrsim 0.21$).
All the cosmologically viable models have an impact of GW propagation \cite{Brax:2015dma}, as shown in Fig. \ref{fig:aT}.
 
Stringent bounds are derived from the constraint on $c_g$ (\ref{eq:bound_speed}). Translated to $\aT$,
\begin{equation}
|\aT| < 9\cdot 10^{-16}\left(\frac{40 \text{Mpc}}{d}\right)\left(\frac{\Delta t}{1.7 \text{s}}\right)\,, \label{eq:cons_aT}
\end{equation}
it implies very strong bounds on $c_4, c_5$. Assuming the non-fine tuned case with no cancellations and noting that $\xi \sim 2$ (range being $1.6 \lesssim \xi \lesssim 3.2 $) we find
\begin{eqnarray}
 |c_4| &<& \frac{\aT}{2\xi^4} \approx 2.8 \cdot 10^{-17} \left(\frac{2}{\xi}\right)^4 \,, \\
 |c_5| &<& \frac{\aT}{0.75 \xi^5} \approx 3.8 \cdot 10^{-17} \left(\frac{2}{\xi}\right)^5 \,
\end{eqnarray}
(compare with cosmology bounds $c_4 = 0.008^{+0.11}_{-0.026}$, $c_5 = -0.013^{+0.023}_{-0.12}$ at 95\% \cite{Renk:2017rzu}). 
This in turn constrains the effective Planck mass and its running to be 
\begin{equation}
 |M_*^2-1| < 1.9\cdot 10^{-15} \,,\qquad 
 |\alpha_M| < 1.9\cdot 10^{-15}\,. \label{eq:cons_aM}
\end{equation}
Note that the bounds on $\Ms$ and $\aM$ (\ref{eq:cons_aM}) are specific to Galileon gravity and will in general be independent from those of $\aT$ in other models.
The most viable Galileon model in this light is a tiny deviation from the cubic Galileon ($c_4=c_5=0$), which is incompatible with the ISW measurements at $7\sigma$ level (Note however that generalizations of the cubic Galileon have been shown to fit ISW data \cite{Kimura:2011td}).

Quintic Galileon models compatible with GW170817 exist on the very narrow and fine-tuned region of the parameter space where $\Delta t \approx \frac{1}{2}\int_{t_E}^{t_O} \alpha_T(t^\prime)dt^\prime \lesssim 1.7$s (Fig. \ref{fig:aT} left).
A second multi-messenger event would, strictly speaking, be necessary to discard this possibility. 
However, such fine-tuning will not be robust to deviations from the cosmological solution, as we discuss next. 

\paragraph{\bf Setting $c_g =1$ on arbitrary backgrounds.}\label{sec:viable_alternatives}

The appearance of an anomalous speed, $\aT\neq0$, can be understood in terms of an effective geometry for the tensor perturbations $\mathcal{G}^{\mu\nu}$, with a different causal structure than the metric field $g^{\mu\nu}$ \cite{Bettoni:2016mij}. 
For Horndeski and beyond Horndeski the form is
\be\label{eq:eff_metric_general}
\mathcal{G}^{\mu\nu}=C g^{\mu\nu} +D\phi^{,\mu} \phi^{,\nu} + E\phi^{;\mu\nu} \,,
\ee
where the coefficients depend on $\phi$ and its derivatives, and all quantities are local. GWs propagation is determined by the on-shell \textit{GW-cone} condition  $\mathcal{G}^{\mu\nu}k_\mu k_\nu = 0$, for $k_\mu = (\omega,\vec k)$, and the propagation speed is $c_g^2(\vec k) =\omega^2(\vec k)/k^2$.
The anomalous GW speed occurs whenever $\mathcal{G}^{\mu\nu} \neq \Omega(x) g^{\mu\nu}$, i.e., $D,E\neq 0$ in (\ref{eq:eff_metric_general}). Quartic theories 
(\ref{eq:L4}) produce $D$-type terms \cite{Bettoni:2016mij}, while quintic theories 
(\ref{eq:L5}) produce also $E$-type terms \cite{Tanahashi:2017kgn}. Both $D,E$ terms can be associated to the presence of the Weyl tensor in the equations of motion \cite{Bettoni:2016mij}.

Satisfying the bound $\aT=0$ requires either both operators leading to $D,E$ to be very suppressed, or an internal cancellation between different terms. But such a cancellation is robust against perturbations only if the different terms involved have the same tensor structure, i.e. different terms contributing to $D$ cancel among themselves and likewise for $E$. 
In contrast, a cancellation between $D$ and $E$ at the level of the cosmological solution is broken by the presence of perturbations. Assuming that such a cancellation exists, computing the effective metric over a perturbed scalar-field solution $\phi = \bar{\phi}(t)+\varphi(x)$ leads to $ \mathcal{G}_{\mu\nu}k^\mu k^\nu = C(\vec k^2 - \omega^2) - 2 E \omega \vec k\cdot \vec\partial \dot\varphi + \cdots$  after boosting so $\phi_\mu = (\omega,\vec 0)$. The GW speed then depends on the direction and can not be compensated
\begin{equation}
 c_g^2 = \frac{\omega^2}{k^2} = 1 + 2\frac{E}{C}  \hat k\cdot \vec\partial \dot\varphi + \cdots\,,
\end{equation}
where the ellipsis denotes terms that modify the GW speed isotropically.
Although this is a second order effect, this deviation will be highly constrained. Thus, tuning the cosmological evolution is not a viable solution to avoid the GWs speed constraint.

\paragraph{\bf Avoiding the GWs speed constraint.}\label{sec:viable_alternatives}

Let us now outline what theories of gravity remain viable late universe models after GW170817. The anomalous GW speed requires two necessary conditions \cite{Bettoni:2016mij}:
(a) non-trivial scalar field configuration that spontaneously breaks Lorentz symmetry and (b) nonzero $D$,$E$ terms in the GW-cone metric (\ref{eq:eff_metric_general}). Note that cosmology ensures field evolution (a), as setting $\aT=0$ via $\dot\phi(t)\approx 0$ cancels any cosmological modified gravity effect altogether. 
Thus, finding viable theories amounts to suppressing or compensating the terms leading to a different causal structure (b).

In the framework of Horndeski the only option is to suppress the terms leading to an anomalous speed. Hence, Horndeski theories are ruled out unless they satisfy
\begin{equation}
G_{4,X} \approx 0 \,,\quad G_5 \approx \text{constant}\,,
\end{equation}
cf. (\ref{eq:L3}-\ref{eq:L5}), with similar restrictions applying to the beyond Horndeski terms introduced in the GLPV theory \cite{Gleyzes:2014dya}. Note that a cancellation of the anomalous speed between $G_4$ and $G_5$ will not be possible in general because they contribute independently to one $D$ and one $E$ term in (\ref{eq:eff_metric_general}). 
The above condition is satisfied only by the simple models contained in $G_2(X,\phi),G_3(X,\phi),G_4(\phi)$.

Viable theories beyond Horndeski can be obtained by modifying the causal structure of the gravitational sector. This can be achieved by applying a disformal transformation of the metric $g_{\mu\nu}\to\tilde g_{\mu\nu}$, where 
\be
 \tilde g_{\mu\nu} = \Omega^2(\phi,X) g_{\mu\nu}+\mathcal{D}(\phi,X)\phi_{,\mu}\phi_{,\nu}\,, \label{eq:disf}
\ee
which changes the GW-cone whenever $\mathcal{D}\neq0$. 
Accordingly, the speed of GWs transforms to%
\footnote{We apply the disformal transformation (\ref{eq:disf}) to the gravity sector only. A field redefinition of the whole action, including matter, will not change the physical ratio $c_g/c$. Note that dependence of the transformation coefficients in $X$ will introduce beyond Horndeski terms in the action (\ref{eq:Li}) \cite{Bettoni:2013diz}.}
\be
\tilde c_g^2=\frac{c_g^2(\tilde X)}{1+2\tilde X \mathcal{D}}\,, \label{eq:disfspeed}
\ee
where $c_g$ is the speed of tensors of the original gravity theory and $-2\tilde X=\tilde g^{\mu\nu}\phi_{,\mu}\phi_{,\nu}$.
This result leaves us with two ways to construct gravity theories with GWs moving at the speed of light: 1) start with a theory with $c_g=1$ and apply a conformal transformation, $\mathcal{D}=0$, or 2) compensate the anomalous speed with a disformal factor, i.e. $\mathcal{D}=(c_g^2-1)/2\tilde X$.

Starting with a $c_g=1$ Horndeski theory and applying a conformal transformation leads to 
\begin{equation}\label{eq:LagConformal}
\Lag_{C} = \frac{1}{16\pi G} \lp \Omega^2 R + 6 \Omega_{,\al}\Omega^{,\al} \rp  +  \tilde\Lag_2 + \tilde\Lag_3 \,,
\end{equation}
with $\Omega=\Omega(X,\phi)$ and where $\tilde\Lag_i$ are the transformed Horndeski $\Lag_2,\Lag_3$ (\ref{eq:L3}) (which transform into combinations of themselves under a disformal relation (\ref{eq:disf})).
The above theory (\ref{eq:LagConformal}), first presented in Ref. \cite{Zumalacarregui:2013pma}, was latter identified as a DHOST theory \cite{Langlois:2015cwa} and hence ghost-free.  It includes mimetic gravity \cite{Chamseddine:2013kea} as a particular case.

Compensating the anomalous speed may also render a theory viable. For a
quartic Horndeski theory (\ref{eq:L4}) with $c_g^2(X)=G_4/(G_4-2XG_{4,X})$ \cite{Bettoni:2016mij}, one needs a beyond Horndeski GLPV Lagrangian \cite{Gleyzes:2014dya}
\be
\begin{split}\label{eq:bh4}
\Lag_{4}^{bH}=F_{4}(\phi,X)\big(\phi_{,\mu}\phi^{;\mu\nu}&\phi_{;\nu\rho}\phi^{\rho}-\phi_{,\mu}\phi^{\mu\nu}\phi_{,\nu}\Box\phi \\
&-X(\left(\Box\phi\right)^{2}-\phi_{;\mu\nu}\phi^{;\mu\nu})\big)\,.
\end{split}
\ee
This term introduces an extra contribution to the speed of gravitational waves that can be used to tune away the anomalous GW speed:
\begin{equation}\label{eq:fine_tuned_H_BH}
c_g^2=\frac{G_4}{G_4-2X(G_{4,X}-XF_4)}=1\Leftrightarrow F_4=G_{4,X}/X\,.
\end{equation}
Not surprisingly, the combined theory is the result of applying a disformal transformation (\ref{eq:disf}), with a suitably chosen $\mathcal{D}$, to the starting Horndeski theory. 
It is important to emphasize that this particular cancellation holds over general backgrounds, as it involves $D$-terms in the effective metric (\ref{eq:eff_metric_general}).  
Our results agree with the independent derivation presented in Ref. \cite{Creminelli:2017sry}.

Thus, the most general ST theory with $c_g=1$ is given by $\Lag_C+\Lag_4+\Lag_4^{bH}$ given by Eqs. (\ref{eq:L4},\ref{eq:LagConformal},\ref{eq:bh4}), subject to the compensation condition (\ref{eq:fine_tuned_H_BH}) (note that the conformal theory contains Horndeski's $G_2,G_3$, and $G_4(\phi)$). This can be understood in the framework of quadratic DHOST theories \cite{Crisostomi:2016czh,deRham:2016wji} for which $c_g^2=G_4/(G_4+2XA_{1})$ (for a cosmological background with a timelike scalar gradient) where $A_1$ is the coefficient of the $\phi_{;\mu\nu}\phi^{;\mu\nu}$ terms in the action. It is very easy to see that this term is canceled by the combination such that the compensation (\ref{eq:fine_tuned_H_BH}) holds. 
Note that, as in Horndeski and GLPV, terms with higher powers of $\nabla\nabla\phi$, cubic DHOST \cite{BenAchour:2016fzp} in this case, cannot help in erasing the anomalous GW speed since they contribute to different terms in the effective metric (\ref{eq:eff_metric_general}).

\begin{figure*}[t!]
\begin{center}
\begin{tikzpicture}
\node (table) {\begin{tabular}{ c | c }
     $c_g=c$ & $c_g\neq c$ \\[2pt] \hline
    \hspace{6cm} & \hspace{6cm}  \\
    General Relativity & quartic/quintic Galileons \cite{Nicolis:2008in,Deffayet:2009wt} \\[2pt]
    quintessence/k-essence \cite{ArmendarizPicon:2000ah} & Fab Four \cite{Charmousis:2011bf} \\[2pt]
    Brans-Dicke/$f(R)$ \cite{Brans:1961sx,Sotiriou:2008rp} & de Sitter Horndeski \cite{Martin-Moruno:2015bda} \\[2pt]
    Kinetic Gravity Braiding \cite{Deffayet:2010qz} & $G_{\mu\nu}\phi^{\mu}\phi^{\nu}$ \cite{Gubitosi:2011sg}, $f(\phi)\cdot$Gauss-Bonnet \cite{Nojiri:2005vv} \\
         &   \\
    Derivative Conformal (\ref{eq:LagConformal}) \cite{Zumalacarregui:2013pma} & quartic/quintic GLPV \cite{Gleyzes:2014dya} \\[2pt]
    Disformal Tuning (\ref{eq:fine_tuned_H_BH})& quadratic DHOST \cite{Langlois:2015cwa} with $A_1\neq0$ \\[2pt]
    quadratic DHOST with $A_1=0$ & cubic DHOST \cite{BenAchour:2016fzp}
\end{tabular}};
\node [rotate=90,blue] at (-6.5,0.27) {Horndeski};
\draw [blue,thick,rounded corners]
  ($(table.south west) !.37! (table.north west)$)
  rectangle 
  ($(table.south east) !.8! (table.north east)$);
 \node [rotate=90,cyan] at (-6.5,-1.7) {beyond H.};
  \draw [cyan,thick,rounded corners]
  ($(table.south west) !.33! (table.north west)$)
  rectangle 
  ($(table.south east) !.02! (table.north east)$);
   \node [midgreen,align=center] at (-3,-2.9) {Viable after GW170817};
   \draw [midgreen,ultra thick,rounded corners, fill=midgreen,fill opacity=0.2]
  ($(-6,-2.6)$) rectangle ($(-0.2,1.7)$);
  \node [red,align=center] at (3,-2.9) {Non-viable after GW170817};
  \draw [red,ultra thick,rounded corners, fill=red,fill opacity=0.2]
  ($(0.2,-2.6)$) rectangle ($(6,1.7)$);
\end{tikzpicture}
\end{center}
\vspace{-15pt}
\caption{Summary of the viable (left) and non-viable (right) scalar-tensor theories after GW170817. Only simple Horndeski theories, $G_{4,X}\approx0$ and $G_5 \approx \text{constant}$, and specific beyond Horndeski models, conformally related to $c_g=1$ Horndeski or disformally tuned, remain viable. 
}
\label{fig:Summary}
\end{figure*}
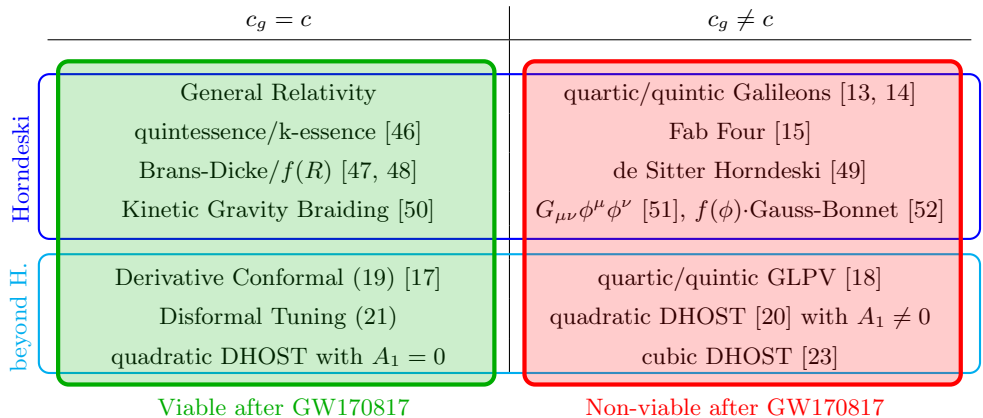

\paragraph{\bf Conclusions.} \label{sec:conclusions}

The coincident arrival of EM and GW signals places one of the strongest bounds available on a large class of scalar-tensor theories that predict an anomalous GW speed. 
The severe constraints on Galileons extends to other scalar-tensor theories: without fine tuning, the quartic and quintic sector of Horndeski, as well as GLPV and several other beyond Horndeski Lagrangians are effectively ruled out as Dark Energy or late universe modifications of gravity. These theoretical classes include some interesting models, such as accelerating solutions due to the weakening of the gravitational force \cite{Lombriser:2015sxa} and self-tuning theories that attempt to solve the cosmological constant problem, and which rely on non-minimal derivative couplings to curvature \cite{Charmousis:2011bf}.

Despite the strong constraints, theories remain that avoid this constraint and thus can still be used to explain DE (see Fig. \ref{fig:Summary}). Within Horndeski's theory these include only the simplest modifications of gravity. 
Beyond Horndeski theory, viable gravities can be obtained in two ways. One can apply a derivative-dependent conformal transformation to those Horndeski models with $c_g=1$, since it does not affect their causal structure.  
Alternatively, one can implement a disformal transformation, which does alter the GW-cone, designed to precisely compensate the original anomalous speed of the theory.

The constraints of GW 170817 extends further into the landscape of gravity theories.
In the case of vector-tensor and scalar-vector-tensor theories, there are several couplings to the curvature that now will be extremely constrained because they modify the speed of GWs, e.g. $R_{\mu\nu}v^{\mu}v^{\nu}$ in vector DE \cite{Jimenez:2008au}. In particular, this test has an impact on Einstein-Aether theories \cite{Jacobson:2004ts}, including some sectors of Ho\v{r}ava gravity \cite{Blas:2014aca}, and more general frameworks such as Generarlized Proca theories \cite{DeFelice:2016yws}. TeVeS \cite{Sagi:2010ei} and MOND-like theories \cite{Chesler:2017khz,Boran:2017rdn} are as well critically affected by this bound. 
Massive gravity \cite{deRham:2010kj}, bigravity \cite{Hassan:2011zd} and multi-gravity \cite{Hinterbichler:2012cn} remain viable as long as the graviton mass is small and matter couples minimally to one of the metrics.

In summary, multi-messenger GW astronomy has proven to be a powerful tool in the quest of the origin of cosmic acceleration and GW170817 sets a landmark in dark energy research. New DE models and theories of gravity will have to satisfy this strong constraint on the GWs speed. 
Future GW-EM detections will be as well determinant for the search of dynamical DE by better constraining the presence of additional polarizations. 

\begin{acknowledgments}
\textbf{Acknowledgements:}
We acknowledge the use of \texttt{hi\_class} \cite{Zumalacarregui:2016pph} to solve the Galileon Cosmology. We are grateful to J. Beltran-Jimenez, D. Bettoni, D. Blas, L. Heisenberg and K. Hinterbichler for useful conversations and G. Horndeski and J. Renk for comments on the manuscript. JME is supported by the Spanish FPU Grant No. FPU14/01618. He thanks UC Berkeley and BCCP for hospitality during his stay there and UAM for financial support. He is also supported by the Spanish Research Agency (Agencia Estatal de Investigación) through the grant IFT Centro de Excelencia Severo Ochoa SEV-2016-0597. MZ is supported by the Marie Sklodowska-Curie Global Fellowship Project NLO-CO. He thanks the PSI$^2$ DarkMod program for support and its participants for useful discussions.
\end{acknowledgments}
 
\textbf{Note added:} Other papers with similar conclusions appeared simultaneously to this work \cite{Creminelli:2017sry,Sakstein:2017xjx,Baker:2017hug}.

\bibliography{gw_refs}

\end{document}